\documentclass[seceq]{ptptex}

\usepackage{graphicx}
\usepackage{colordvi}

\title{
Quantum chaos in supersymmetric QCD at finite density
}

\author{
Elmar \textsc{Bittner}$^1,$ Simon \textsc{Hands}$^2,$ Harald \textsc{Markum}$^3,$ and Rainer \textsc{Pullirsch}$^3$
}
\inst{
$^1$Institut f\"ur Theoretische Physik, Universit\"at Leipzig, Augustusplatz 10/11, D-04109 Leipzig, Germany\\
$^2$Department of Physics, University of Wales Swansea, Singleton Park, 
\\Swansea SA2 8PP, UK\\
$^3$Atominstitut, Technische Universit\"at Wien, Wiedner Hauptstra\ss e 8-10/141, A-1040 Wien, Austria
}

\abst{
  We investigate the distribution of the spacings of adjacent
  eigenvalues of the lattice Dirac operator.  At zero chemical
  potential $\mu$, the nearest-neighbor spacing distribution $P(s)$
  follows the Wigner surmise of random matrix theory both in the
  confinement and in the deconfinement phase.  This is indicative of
  quantum chaos.  At nonzero chemical potential, the eigenvalues of
  the Dirac operator become complex and we discuss how $P(s)$ can be
  defined in the complex plane.  Numerical results from an SU(2)
  simulation with staggered fermions in fundamental and adjoint
  representations are compared with predictions
  from non-hermitian random matrix theory, and agreement with the
  Ginibre ensemble is found for $\mu\approx 0.5$.
}

\begin{document}

\maketitle

\section{Introduction}
\label{sec1}

The spectrum of the QCD Dirac operator, both in the continuum and on
the lattice, has several universal properties.  By ``universal'' one
means ``independent of the details of the dynamics'', e.g., independent
of the precise values of the simulation parameters on the lattice.
Such universal features can be described by random matrix theory
(RMT).  In this contribution, we are concerned with the eigenvalue
fluctuations in the bulk of the spectrum on the scale of the local
mean level spacing, measured by the distribution $P(s)$ of
the spacings $s$ of adjacent eigenvalues.  We will consider gauge
group SU(2) and staggered fermions in the fundamental representation
which are related to the chiral symplectic
ensemble of RMT.  At zero chemical potential $\mu$, all Dirac
eigenvalues are purely imaginary, and it has been shown in lattice
simulations that $P(s)$ agrees with the Wigner surmise of RMT,
\begin{equation}
  \label{eq1}
  P_{\rm W}(s)=\frac{262144}{729\pi^3}\,s^4\,e^{-\frac{64}{9\pi}s^2}\:,
\end{equation}
both in the confinement and in the deconfinement phase \cite{Hala95}.
This result implies that the Dirac eigenvalues are strongly
correlated, and is indicative of quantum chaos, according to the
conjecture by Bohigas, Giannoni, and Schmit \cite{Bohi84b}.  In
contrast, quantum systems with uncorrelated eigenvalues, corresponding
to classically integrable systems, obey a Poisson distribution, 
\begin{equation}
P_{\rm P}(s)=e^{-s} \: .
\end{equation}
Further, we present a first
analysis for gauge group SU(2) with staggered fermions in the adjoint
representation being related to the chiral orthogonal ensemble with
the Wigner surmise
\begin{equation}
  \label{eq1a}
  P_{\rm W}(s)=\frac{\pi}{2}\,s\,e^{-\frac{\pi}{4}s^2}\:.
\end{equation}
While fermions transforming with respect to the fundamental representation
describe quarks, fermions transforming according to the adjoint representation
can be interpreted as gluinos. Thus the latter case with
\begin{equation}
  \label{eq1b}
   U^{\rm{adj}}_{vu,x\mu} = \frac{1}{2} {\rm Tr} \left( U^{\dagger}_{x\mu}
   \tau_v U_{x\mu} \tau_u \right)
\end{equation}
constitutes a lattice version of supersymmetric Yang-Mills theory~\cite{Gerst}.

We focus on nonzero chemical potential, $\mu\ne0$, where
the lattice Dirac matrix generalizes to
\begin{eqnarray}
  \label{eq2}
  M_{x,y}(U,\mu)&=&
  \frac{1}{2a} \sum\limits_{\nu=\hat{x},\hat{y},\hat{z}}
  \left[U_{\nu}(x)\eta_{\nu}(x)\delta_{y,x\!+\!\nu}-{\rm h.c.}\right]\nonumber \\ 
  &+&
  \frac{1}{2a}\left[U_{\hat{t}}(x)\eta_{\hat{t}}(x)e^{\mu}
    \delta_{y,x\!+\!\hat{t}}
    -U_{\hat{t}}^{\dagger}(y)\eta_{\hat{t}}(y)
    e^{-\mu}\delta_{y,x\!-\!\hat{t}}\right]\:,
\end{eqnarray}
with the Kawamoto-Smit phases $\eta$.  The eigenvalues of this matrix
are complex.  In this case, $P(s)$ represents the spacing distribution
of nearest neighbors in the complex plane. For each eigenvalue
$z_0$ one has to find the eigenvalue $z_1$ for which $s=|z_1-z_0|$ is
smallest, with a subsequent average over $z_0$.  This definition
assumes that the spectral density is constant over a bounded region in
the complex plane (and zero outside).  Generally, this is not the case
so that an unfolding procedure must be applied; see Sec.~\ref{sec2}.
However, the spectral density of the so-called Ginibre ensemble of RMT,
where real and imaginary parts of the eigenvalues have the same
average size, is constant inside a circle and zero outside,
respectively \cite{Gini65}.  In this case, $P(s)$ is given by
\cite{Grob88}
\begin{equation}
  \label{eq3}
  P_{\rm G}(s)=c \, p(cs)\:,
\end{equation}
with
\begin{displaymath}
  \label{eq4}
  p(s)=2s\lim_{N\to\infty}\left[\prod_{n=1}^{N-1}e_n(s^2)\,e^{-s^2}\right]
  \sum_{n=1}^{N-1}\frac{s^{2n}}{n!e_n(s^2)}\:,
\end{displaymath}
$e_n(x)=\sum_{m=0}^n x^m/m!$, and $c=\int_0^\infty ds \, s \,
p(s)=1.1429...$\\  In contrast, the Poisson distribution in the complex
plane, representing uncorrelated eigenvalues, becomes
\begin{equation}
  \label{eq5}
  P_{\bar{\rm P}}(s)=\frac{\pi}{2}\,s\,e^{-\frac{\pi}{4}s^2}\:.
\end{equation}
This should not be confused with the Wigner distributions (\ref{eq1}) and (\ref{eq1a}).
In the following, we will study the Dirac spectrum on the lattice at
various values of $\mu\neq 0$ and compare the resulting $P(s)$ with
Eqs.~(\ref{eq3}) and (\ref{eq5}).

\section{Analysis of complex spectra}
\label{sec2}

\begin{figure}[hp]
\hspace*{-10mm}\includegraphics[width=16cm]{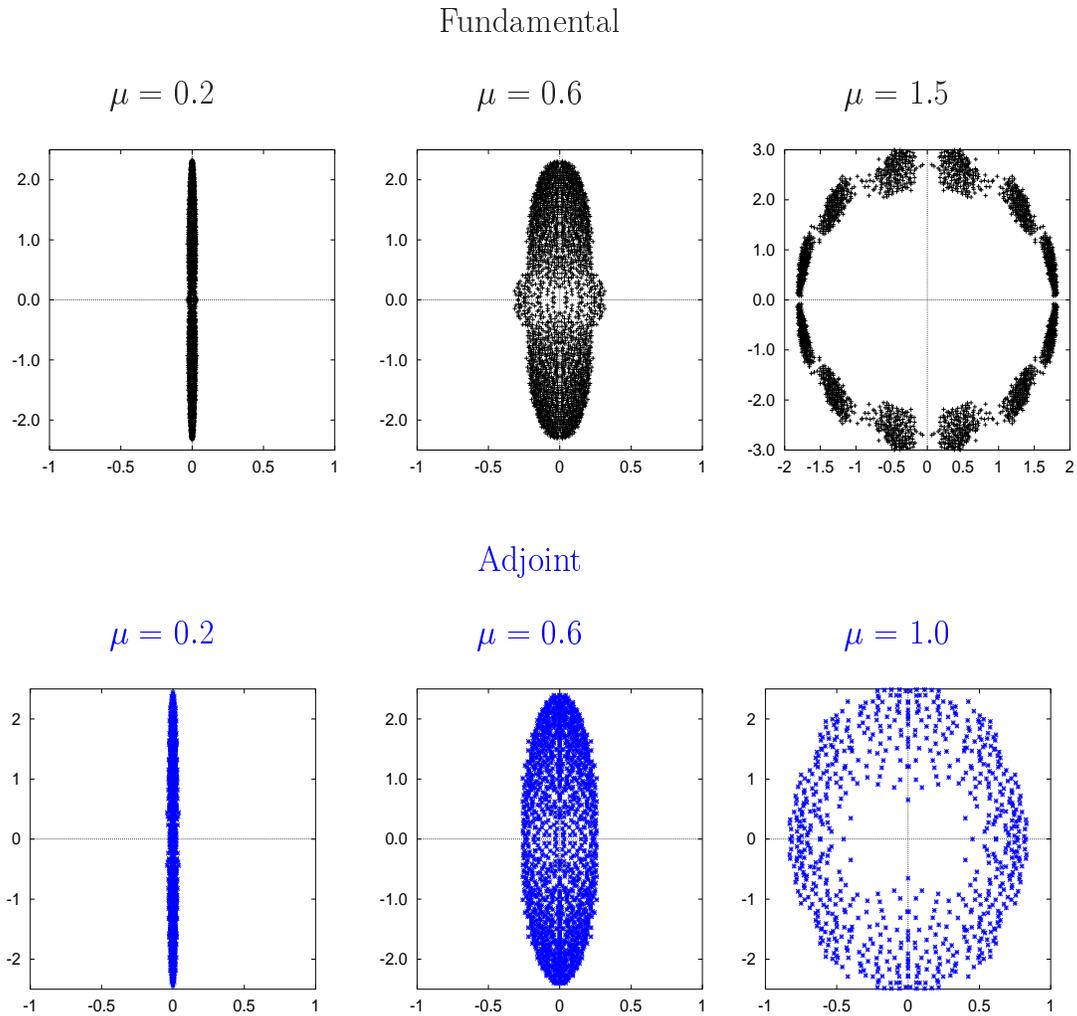}
\caption{
    Complex eigenvalues of the Dirac operator at various values
    of $\mu$ for a typical equilibrium configuration of two-color QCD with staggered
    fermions in the fundamental and adjoint representation (horizontal axes = real
    parts, vertical axes = imaginary parts, in units of $1/a$).
\label{fig1}
 }
\end{figure}

We report on simulations done with gauge group SU(2) in the confinement
region; specifically we studied a $6^4$ system at $\beta=1.3$ with 2 flavors of staggered
fermion with bare mass $ma=0.07$ in the fundamental representation~\cite{india}, and
a $4^4$ system at $\beta=2.0$ with 1 flavor of staggered fermion with $ma=0.1$ in the 
adjoint representation (corresponding respectively to $N_f=8$ and 4 continuum
flavors).
For these systems the fermion determinant is real and lattice simulations become
feasible~\cite{Hands}. We sampled a few 100 -- 1000 independent configurations
for each $\mu$-value.

The eigenvalue spectrum is shown in Fig.~\ref{fig1} for three different
values of $\mu$.  The size of the real parts of the eigenvalues grows
with $\mu$ as expected.  While the spectrum is confined to a bounded
region in the complex plane, the spectral density is certainly not
constant.  Therefore, the spectrum has to be unfolded.  In one
dimension, unfolding is a local rescaling of the eigenvalue density
such that the density on the unfolded scale is equal to unity.  To 
unfold spectra in the entire complex plane we proceed as follows.  The
spectral density has an average and a fluctuating part,
$\rho(x,y)=\rho_{\rm av}(x,y)+\rho_{\rm fl}(x,y)$.  Unfolding means to
find a map $z'=x'+iy'=u(x,y)+iv(x,y)$ such that $\rho_{\rm
  av}(x',y')\equiv 1$.  Since the probability has to be invariant,
$\rho_{\rm av}(x',y')dx'dy' = dx'dy' = \rho_{\rm av}(x,y)dxdy$, we
find that $\rho_{\rm av}(x,y)$ is the Jacobian of the transformation
from $(x,y)$ to $(x',y')$,
$  \label{eq6}
  \rho_{\rm av}(x,y)=\left|\partial_xu\,\partial_yv-
    \partial_yu\,\partial_xv\right|\:.
$
Choosing $y'=v(x,y)=y$ yields $\rho_{\rm av}(x,y)=|\partial_xu|$ and,
thus, 
$
  \label{eq7}
  x'=u(x,y)=\int_{-\infty}^xdt\rho_{\rm av}(t,y)\:.
$
Essentially, this is a one-dimensional unfolding in strips parallel to
the real axis~\cite{Fyod97}. For a fixed bin in $y$, $\rho_{\rm av}(x,y)$ is
determined by fitting $\rho(x,y)$ to a low-order polynomial.  $P(s)$
is then constructed from the constant unfolded density as explained in
Sec.~\ref{sec1}, normalized such that $\int_0^\infty ds\,s\,P(s)=1$.

Several remarks are in order.  (i) We have checked spectral
ergodicity.  If only parts of the spectral support are considered, the
results for $P(s)$ do not change.  (ii) If the spectral density has
``holes'', we split the
spectral support into several convex pieces and unfold them
separately.  This is justified by spectral ergodicity.  (iii)
Unfolding each spectrum separately and ensemble unfolding yield
the same results for $P(s)$.  (iv) The results for $P(s)$ are stable
under variations of the degree of the fit polynomial and of the bin
sizes in $x$ and $y$.

\section{Results and discussion}
\label{sec3}

\begin{figure}[hp]
\hspace*{-10mm}\includegraphics[width=16cm]{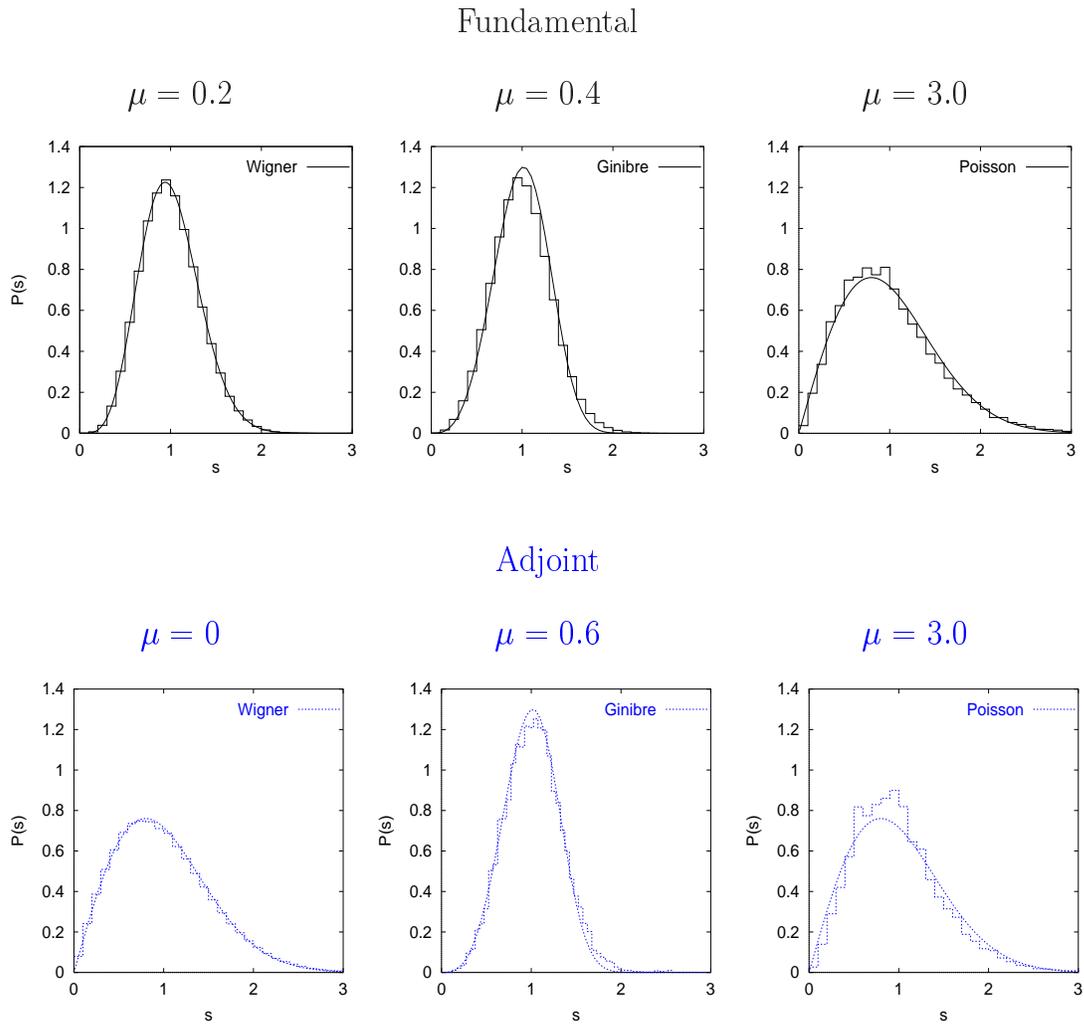}
\caption{
    Nearest-neighbor spacing distribution of the Dirac matrix
    with fundamental and adjoint fermions for various values of $\mu$.  The
    histograms represent the lattice data. The curves corresponding 
    to the Wigner, Ginibre and Poisson distribution are inserted.
\label{fig2}
 }
\end{figure}

Our results for $P(s)$ are presented in Fig.~\ref{fig2}. As a
function of $\mu$, we expect to find a transition from 
Wigner to Ginibre behavior in $P(s)$. This was clearly seen in 
color-SU(3) with $N_f = 3$ flavors and quenched chemical
potential~\cite{Mark}, where differences between both curves are
more pronounced. For the symplectic ensemble of color-SU(2) with
staggered fermions in the fundamental representation, the
Wigner and Ginibre distributions are very close to each other and
thus harder to distinguish, but they are reproduced for $\mu=0$
and $\mu=0.4$, respectively. For the orthogonal ensemble with
staggered fermions in the adjoint representation this transition
is more drastic.

Increasing $\mu > 1.0$, the lattice results for $P(s)$ deviate substantially
from the Ginibre distribution and can be interpreted as Poisson
behavior, corresponding to uncorrelated eigenvalues. (In the Hermitian
case at nonzero temperature, lattice simulations only show a transition
to Poisson behavior for $\beta \to \infty$ when the physical
box size shrinks and the theory becomes free~\cite{Hala95}.)
A plausible explanation of the transition to Poisson behavior is provided
by the following two (related) observations. First, for large $\mu$
the terms containing $e^{\mu}$ in Eq.(\ref{eq2}) dominate the Dirac
matrix, giving rise to uncorrelated eigenvalues. Second, for large
$\mu$ the fermion density on the finite lattice reaches saturation
due to the limited box size and the Pauli exclusion principle.

In conclusion, we have investigated the nearest-neighbor spacing
distribution of the lattice Dirac operator for two-color QCD with
fermions in the fundamental and adjoint representation switching on
a finite chemical potential. We find a transition from a Wigner to a Ginibre
distribution around the phase transition for both types of fermions.
This means that quantum chaos persists deep into the high-density
phase.

\section*{Acknowledgements}
We thank Maria-Paola Lombardo for previous collaboration within fermions
in the fundamental representation. This work was partly supported by 
FWF-Project P14435-TPH ``Random Matrix Theory and Quantum Chaos in Quantum Field Theories'' 
and by the EU-Network HPRN-CT-1999-000161
``Discrete Random Geometries: From Solid State Physics to Quantum Gravity''.

\end{document}